\documentclass{aip-cp}

\usepackage[numbers]{natbib}
\usepackage{rotating}
\usepackage{graphicx}
\usepackage{url}
\usepackage{multirow}
\usepackage{booktabs}

\begin{document}

\title{Strange Nuclear Physics from QCD on Lattice}

\author[aff1,aff2]{Takashi Inoue\corref{cor1}}
\author{HAL QCD Collaboration}

\affil[aff1]{Nihon University, College of Bioresource Sciences, Fujisawa 252-0880, Japan}
\affil[aff2]{ Theoretical Research Division, Nishina Center, RIKEN, Wako 351-0198, Japan}
\corresp[cor1]{Corresponding author: inoue.takashi@nihon-u.ac.jp}

\maketitle

\def \etal{{\it et al.\,}}
\def \etc{{\it etc.\,}}
\def \ie{{\it i.e.\,}}
\def \eg{{\it e.g.\,}}
\def \brav #1|{\mbox{$\langle {#1}|$}}
\def \ketv #1>{\mbox{$|{#1}\rangle$}} 
\def \bracket<#1>{\mbox{$\langle {#1}\rangle$}}
\def \mate<#1|#2|#3>{\mbox{$\langle {#1}|\,{#2}\,|{#3}\rangle$}}

\begin{abstract}
We study single-particle potential of $\Lambda$, $\Sigma$, and $\Xi$ hyperons in nucleonic matter
starting from the fundamental theory of the strong interaction, QCD.
First we carry out a lattice QCD numerical simulation, and extract baryon-baryon interactions from QCD by means of the HAL QCD method. 
We employ a full QCD gauge configuration ensemble at almost physical point so that we can study the physical world,
hence mass of hadrons are nearly physical, for example, pion mass is 146 MeV, kaon mass is 525 MeV, and nucleon mass is 958 MeV.
Then, we apply the obtained hyperon interactions to the Brueckner-Hartree-Fock many-nucleon theory,
and calculate single-particle potential of hyperons in nucleonic matter $U_{Y}(k)$.
We obtain for hyperons stopping in the symmetric nuclear matter at the normal nuclear matter density,
$U_{\Lambda}(0)=-28$~MeV, $U_{\Sigma}(0)=+15$~MeV, and  $U_{\Xi}(0)=-4$~MeV
with a statistical error about $\pm 2$ MeV associated with our Monte Carlo simulation.
These results are qualitatively compatible with values suggested from hypernuclear experiments.
This success is remarkable and very encouraging since this proves that our approach to strange nuclear physics
starting from QCD is essentially correct. 
\end{abstract}

\section{INTRODUCTION}

In this decades, $\Lambda$, $\Sigma$ and $\Xi$ hyperons are serious subject in physics of neutron stars.
It has been naturally considered that hyperons ($Y$) exist in core of neutron stars.
Existence of hyperon has a significant impact on structure, evolution, and property of neutron stars.
However, equation of state of baryonic matter including hyperons would be considerably soft,
and it seems incompatible with recent discovery of heavy neutron stars~\cite{Demorest:2010bx}.
This is the so called hyperon puzzle in neutron stars, and one of the most challenging problem in modern physics.

Hyperon emergence in matter is determined by its chemical potential $\mu_Y(\rho$),
and $\mu_Y(\rho)$ strongly depends on hyperon interaction through single-particle potential of hyperon in matter $U_Y(k)$.
Therefore, hyperon interaction and corresponding $U_Y(k)$ are essentially important to solve the hyperon puzzle.

As you know, hyperon-nucleon ($YN$) scattering experiment is difficult because of weak decay of hyperon,
and hence experimental information of hyperon interaction have been obtained mainly from study of hypernuclei. 
By using rich data of $\Lambda$-hypernuclei, single-particle potential of $\Lambda$ in nuclei is well determined,
and $\Lambda N$ interaction is revealed relatively well.
While, experimental data of $\Sigma$-nucleus system and $\Xi$-nucleus system are limited.
A lot of effort has been devoted to measure these system and is continuing presently at many facilities.
For example, the first evidence of $\Xi$-hypernucleus, ${}^{14}_{~\Xi}\mbox{N}$, was found recently at KEK in Japan~\cite{Nakazawa:2015joa}.

Hyperon interaction have been studied theoretically for many years.
Several useful potentials of hyperon interaction had been made by basing some model, \eg meson exchange models, quark models, and so on.
These potentials have been applied to investigate strange nuclear physics,
including theoretical prediction of $U_Y(k)$ in matter~\cite{Baldo:1999rq,Kohno:2009vk,Yamamoto:2014jga}.
Since models are more or less phenomenological, some uncertainties exist
in the model hyperon potentials inevitably due to lack of experimental data.

In 2006, a brand-new method was proposed to extract nucleon-nucleon interaction from QCD on lattice~\cite{Ishii:2006ec}.
This so called HAL QCD method has been applied successfully to hyperon-nucleon systems~\cite{Nemura:2008sp},
general baryon-baryon systems~\cite{Inoue:2010hs}, meson-baryon systems~\cite{Ikeda:2010sg} and so on.
This means that today we are able to obtain hyperon interactions without requiring experimental data but relying on QCD.
It is very interesting to see what QCD induced hyperon interaction predicts about hyperons in nucleonic matter.
Hence, in this paper, first we derive hyperon interactions in lattice QCD simulation,
and then apply resulting interaction potentials to study hyperons in nucleonic matter~\cite{Inoue:2016qxt}.

\section{Method to extract hyperon interactions from QCD on lattice}
In this section, we briefly describe a way to extract hyperon interactions from QCD on lattice.
For a concrete example, let us consider interaction in two-baryon sector with strangeness $S$=$-1$, isospin $I$=$1/2$, and ${}^{1}S_0$ partial wave.
In this sector, there are two flavor eigenstates $\ketv i=1,2>$, 
which are nothing but flavor irreducible representations in flavor {\it SU}(3) symmetric case, 
but in general exist also in flavor {\it SU}(3) broken case. 
Each eigenstate $\ketv i>$ has corresponding vector of wave function in two-baryon channel basis $\{ \Lambda N, \Sigma N \}$,
which we denote as $( \psi^{(i)}_{\Lambda N}, \psi^{(i)}_{\Sigma N} )^t$.
These wave functions should obey the Schr{\" o}dinger equation, in the Euclidean space-time, 
\begin{equation}
\label{eqn:scheq}
\left( 
 \begin{array}{ccc}
 -\frac{\partial}{\partial t} &         0       \\
        0       &  -\frac{\partial}{\partial t}
 \end{array}
\right)
\left( 
 \begin{array}{c}
 \psi^{(i)}_{\Lambda N} \\
 \psi^{(i)}_{\Sigma  N}
 \end{array}
\right)
=
\left[
\left( 
 \begin{array}{cc}
     M_{\Lambda N} + \frac{-\nabla^2}{2\mu_{\Lambda N}} &  0 \\
 0 & M_{\Sigma  N} + \frac{-\nabla^2}{2\mu_{\Sigma  N}}
 \end{array}
\right)
+
\left( 
 \begin{array}{ll}
 U_{\Lambda N,\Lambda N} & U_{\Lambda N,\Sigma N} \\
 U_{\Sigma  N,\Lambda N} & U_{\Sigma  N,\Sigma N} 
 \end{array}
\right)
\right]
\left( 
 \begin{array}{c}
 \psi^{(i)}_{\Lambda N} \\
 \psi^{(i)}_{\Sigma  N}
 \end{array}
\right)
\end{equation}
where we've introduced a matrix of interaction potential in the coupled channel space $U_{a b}$, which we are interested in, 
and $M_{B B'}$, $\mu_{B B'}$ are the two-baryon total and reduced mass, respectively.
The potentials $U_{a b}$ are non-local but energy independent in our definition.
Since the potentials are identical for the two eigenstates, they can be obtained by inverting a set of the Schr{\" o}dinger equation at each point as
\begin{equation}
\label{eqn:pot1}
\left( 
 \begin{array}{ll}
 U_{\Lambda N,\Lambda \Lambda} & U_{\Lambda N,\Sigma N} \\
 U_{\Sigma  N,\Lambda \Lambda} & U_{\Sigma  N,\Sigma N} 
 \end{array}
\right)
 =
\left[
 \left( 
 \begin{array}{cc}
          \frac{\nabla^2}{2\mu_{\Lambda N}} - \frac{\partial}{\partial t} - M_{\Lambda N} & 0 \\
  0 &     \frac{\nabla^2}{2\mu_{\Sigma  N}} - \frac{\partial}{\partial t} - M_{\Sigma  N}
 \end{array}
 \right)
 \left( 
 \begin{array}{ccc}
 \psi^{(1)}_{\Lambda N} & \psi^{(2)}_{\Lambda N} \\
 \psi^{(1)}_{\Sigma  N} & \psi^{(2)}_{\Sigma  N}
 \end{array}
 \right)
\right]
\left( 
 \begin{array}{cc}
 \psi^{(1)}_{\Lambda N} & \psi^{(2)}_{\Lambda N} \\
 \psi^{(1)}_{\Sigma  N} & \psi^{(2)}_{\Sigma  N}
 \end{array}
\right)^{-1}
\end{equation}
where the derivative operators act inside the square bracket.
We've wrote as if $U_{ab}$ are local because we expand $U_{ab}$ in terms of local functions with the derivative operator,
and the leading term $V_{ab}$ is just a local potential.

While, in lattice QCD simulation, we calculate the so called 4-point correlation functions
\begin{equation}
\label{eqn:4pt}
 \phi_{(BB')(J)}(\vec{r}, t) \equiv \frac{1}{\sqrt{Z_{B} Z_{B'}}} \sum_{\vec{x}}
                         \, \langle 0 \vert B(\vec x + \vec r,t) B'(\vec x,t) {J}(t_0)\vert 0 \rangle
\end{equation}
where $B(\vec y,t)B'(\vec x,t)$ is a product of baryon field operators at sink,
${J}(t_0)$ is a source operator which creates two baryons at $t_0$,
and $Z_{B}$ is a renormalization factor of the baryon field.
Because we can use two kinds of baryon operator pair, $\Lambda N$ and $\Sigma N$, at both sink and source,
we have four kinds of 4-point correlation functions $\phi_{a b}(\vec r,t)$ in this sector.
It is known that 4-point correlation function contains scattering observables, which are information of interaction,
exactly same way as quantum mechanical wave function does~\cite{Ishizuka:2009bx,Aoki:2012bb}.
The flavor eigenstate $\ketv i>$ can be exclusively generated by a source with a particular linear combination of the operator pairs,
let's say $J^{(i)}(t_0)=c^{(i)}_{\Lambda N} \overline{\Lambda N} + c^{(i)}_{\Sigma N} \overline{\Sigma N}$.
Consequently, the wave function of each eigenstate 
is given by a linear combination of the 4-point correlation functions as
\begin{equation}
\label{eqn:eigen}
\left( 
 \begin{array}{c}
 \psi^{(i)}_{\Lambda N} \\
 \psi^{(i)}_{\Sigma  N}
 \end{array}
\right)
 = 
\left( 
 \begin{array}{cc}
 \phi_{(\Lambda N)(\Lambda N)} & \phi_{(\Lambda N)(\Sigma N)} \\
 \phi_{(\Sigma  N)(\Lambda N)} & \phi_{(\Sigma  N)(\Sigma N)}
 \end{array}
\right)
\left( 
 \begin{array}{c}
 c^{(i)}_{\Lambda N} \\
 c^{(i)}_{\Sigma  N}
 \end{array}
\right) ~.
\end{equation}
By inserting this expression to eq.(\ref{eqn:pot1}), we arrive the formula which we use
to derive hadron interaction potentials from QCD on lattice, in the leading order,
\begin{eqnarray}
\label{eqn:fomula}
\left( 
 \begin{array}{ll}
 V_{\Lambda N,\Lambda N} & V_{\Lambda N,\Sigma N} \\
 V_{\Sigma  N,\Lambda N} & V_{\Sigma  N,\Sigma N} 
 \end{array}
\right)
&=&
\left( 
 \begin{array}{cc}
 \frac{\nabla^2}{2\mu_{\Lambda N}} \phi_{(\Lambda N)(\Lambda N)} &
 \frac{\nabla^2}{2\mu_{\Lambda N}} \phi_{(\Lambda N)(\Sigma  N)} \\
 \frac{\nabla^2}{2\mu_{\Sigma  N}} \phi_{(\Sigma  N)(\Lambda N)} &
 \frac{\nabla^2}{2\mu_{\Sigma  N}} \phi_{(\Sigma  N)(\Sigma  N)}
 \end{array}
\right)
\left( 
 \begin{array}{cc}
 \phi_{(\Lambda N)(\Lambda N)} & \phi_{(\Lambda N)(\Sigma N)} \\
 \phi_{(\Sigma  N)(\Lambda N)} & \phi_{(\Sigma  N)(\Sigma N)} 
 \end{array}
\right)^{-1}
\\
&&+
\left(
 \begin{array}{cc}
 \left(-\frac{\partial}{\partial t} - M_{\Lambda N} \right) \phi_{(\Lambda N)(\Lambda N)} &
 \left(-\frac{\partial}{\partial t} - M_{\Lambda N} \right) \phi_{(\Lambda N)(\Sigma  N)}  \\
 \left(-\frac{\partial}{\partial t} - M_{\Sigma  N} \right) \phi_{(\Sigma  N)(\Lambda N)} &
 \left(-\frac{\partial}{\partial t} - M_{\Sigma  N} \right) \phi_{(\Sigma  N)(\Sigma  N)}
 \end{array}
\right)
\left( 
 \begin{array}{cc}
 \phi_{(\Lambda N)(\Lambda N)} & \phi_{(\Lambda N)(\Sigma N)} \\
 \phi_{(\Sigma  N)(\Lambda N)} & \phi_{(\Sigma  N)(\Sigma N)} 
 \end{array}
\right)^{-1}
\nonumber
\end{eqnarray}
where the leading order potentials are divided into two terms, namely a Laplacian part and a time-derivative part, for convenience in analysis.
Note that the coefficients $(c^{(i)}_{\Lambda N}, c^{(i)}_{\Sigma N})^t$ vanish in this formula.
This means that we do not need to obtain eigenstates and we can derive the potentials directly from data of the 4-point functions. 
In order to apply this formula, linear independence of vector
$(\phi_{(\Lambda N)(\Lambda N)},\phi_{(\Sigma  N)(\Lambda N)})^t$ and
$(\phi_{(\Lambda N)(\Sigma  N)},\phi_{(\Sigma  N)(\Sigma  N)})^t$ is necessary.
If the sink-source time-separation, $t-t_0$ in eq(\ref{eqn:4pt}), is extremely large,
both the vectors become a wave function of the ground state of the sector and they are linearly dependent on each other.
However, since we always use a moderate size of $t-t_0$ in our calculations as you will see, this will never be a problem.

Thus, we can derive coupled channel potentials of hyperon interactions in lattice QCD numerical simulation. 
Extension to other sector is straightforward. For example, in $S$=$-2$, $I$=$0$, ${}^{1}S_0$ two-baryon sector,
we obtain a $3\times 3$ matrix of potential in $\{\Lambda \Lambda, \Xi N, \Sigma \Sigma \}$ coupled channel space.

In this HAL QCD method, we do NOT need to separate or suppress two-body excitation in the 4-point functions.
This is a grate advantage of this method over the conventional direct one in which
separation or suppression of two-body excitation is crucial but difficult or impossible
to achieve for multi-baryon systems~\cite{Iritani:2016jie}.
In the HAL QCD method, we just need to suppress contamination of excited single hadron,
and we can achieve it by taking a certain size of the sink-source separation $t-t_0$.
Because the signal-over-noise ratio become worse rapidly at large separation,
we normally take $t-t_0 = 1 \sim 2$ fm depending on hadrons involved and quality of data of the 4-point function.

\section{Setup of lattice QCD simulation}
\label{sec:setup}

\begin{table}[t]
\caption{Mass of the pseudo-scalar mesons and the octet baryons measured in the lattice QCD simulation with the K-configuration set.
The mass of pion and kaon are taken from ref.~\cite{Ishikawa:2015rho}}
\label{tbl:mass}
\begin{tabular}{ccccccc}
 \toprule
  Hadron     &  $\pi$   &   K   &   N   & $\Lambda$ & $\Sigma$ & $\Xi$ \\
  \midrule
  Mass [MeV] &   146    &  525  &  958(3)  &   1140(2)   &   1223(2)  & 1354(1)  \\ 
 \bottomrule
\end{tabular}
\end{table}

In general, we need an ensemble of gauge configuration to cay out lattice QCD numerical simulation.
Recently, a configuration set is generated on the K computer at RIKEN AICS in Japan,
by a collaboration in HPCI Strategic Program field 5 project 1.
There, the stout smeared Wilson clover action for quarks and the Iwasaki gauge action for gluon are employed. 
Details of the K-configuration set can be found in ref.~\cite{Ishikawa:2015rho}.
We employ this configuration set because it is suitable for our purpose~\cite{Doi:2017zov}.
First, it is generated at almost physical point, which enable us to study the real world.
Table~\ref{tbl:mass} list mass of hadrons measured in lattice QCD simulation with the K-configuration set.
One sees that measured hadron masses almost agree with the physical ones.
Second, the spatial volume of this configuration set is large $V \simeq (8.1~\mbox{fm})^3$,
which is very important for us to derive baryon-baryon interactions.

To evaluate the 4-point correlation functions $\phi(\vec r,t)$ in eq.(\ref{eqn:4pt}),
we adopt the wall type quark source $J(t_0)$ and the point type baryon field operator $B$ and $B'$ at sink.
We choose the wall type source just for convenience. We can use any type of source.
We choose the point type operator at sink so that the non-locality of potential $U_{a b}(\vec r, \vec r\,')$ is minimized, 
which is our scheme to define potentials.
We treat the minimized non-locality of $U_{a b}(\vec r, \vec r\,')$ by means of the derivative expansion and truncation.
We know that convergence of the expansion is fast owing to our choice of operator at sink~\cite{Kawai:2017goq},
and hence we maintain only the leading order term $V_{a b}(\vec r)$ in this study.
Refinement by taking the next leading order term of the expansion is one of our future challenges.

We utilize all 414 gauge configurations available in the set.
To avoid a wraparound artifact, we put the Dirichlet boundary conditions in the temporal direction 
and take the average over forward and backward propagations in time.
In order to reduce noise and enhance signal, we repeat measurement $4 \times 96$ times for each configuration
by rotating axes of lattice and shifting the source time $t_0$, and take the average of them.
Then, we put the obtained noise reduced 4-point function into correspondences of eq.(\ref{eqn:fomula}),
and obtain interaction potentials $V_{ab}(\vec r)$ at last.

\section{Hyperon interaction potentials from QCD}

\begin{figure}[t]
\centering
\includegraphics[width=0.79\textwidth]{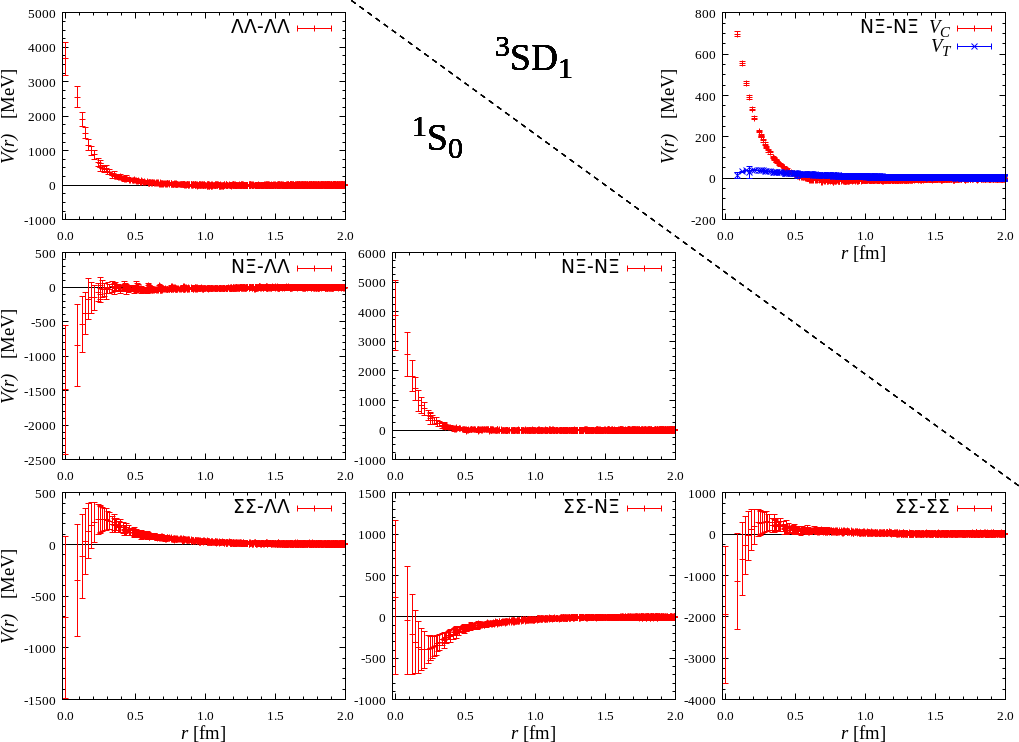}
\caption{Baryon-baryon interaction potentials from QCD in strangeness $S$=$-2$ and isospin $I$=$0$ sector.}
\label{fig:isospin0}
\end{figure} 
\begin{figure}[t]
\centering
\includegraphics[width=0.79\textwidth]{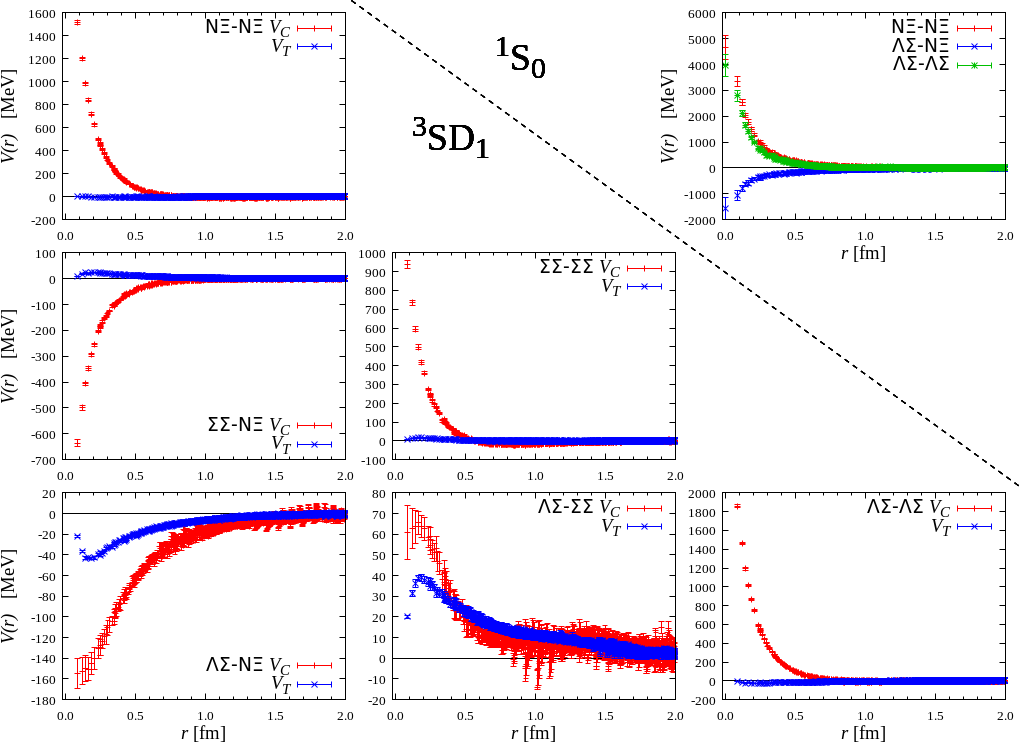}
\caption{Baryon-baryon interaction potentials from QCD in strangeness $S$=$-2$ and isospin $I$=$1$ sector.}
\label{fig:isospin1}
\end{figure} 

We show some of obtained baryon-baryon interaction potentials.
Fig.~\ref{fig:isospin0} shows potentials in strangeness $S$=$-2$ and isospin $I$=$0$ sector,
while Fig.~\ref{fig:isospin1} shows ones in $S$=$-2$, $I$=$1$ sector.
There is also one $I$=$2$, $^1S_0$, $\Sigma\Sigma$ interaction in $S$=$-2$ sector, but it is suppressed.
For $^1S_0$ partial waves, potentials are just central $V(r)$.
While, for $^3S_1$-$^3D_1$ coupled partial waves, the central potential $V_C(r)$ and the tensor potential $V_T(r)$ are given.
The vertical bars stand for statistical error associated with our Monte Carlo calculation, estimated with the Jackknife method.
These potentials are obtained by using the 4-point function at a sink-source separation $t-t_0=12$ in lattice unit $a\simeq0.0846$ fm.
We fix the separation $t-t_0=12$ throughout this paper, since we find, by changing the separation,
that the resulting potentials are stable between $t-t_0=11 \sim 13$,
{\it i.e.} agree with each other within the error bar, and the choice $t-t_0=12$ is most convenient with our data set.
We have obtained also potentials of S-wave baryon-baryon interactions in $S$=$0$,$-1$,$-3$, and $-4$ sectors.
But, we do not show them in this paper because they will be presented in separated papers.
For example, one can find potentials in $S$=$-1$ sectors in ref.~\cite{Nemura:2018}.

Because there are many potentials between many channels in Fig.~\ref{fig:isospin0} and Fig.~\ref{fig:isospin1},
it is difficult to grasp feature of baryon-baryon interaction in $S$=$-2$ sector.
As you know, if the flavor {\it SU}(3) symmetry holds exactly, all octet-baryon pairs can be classified into six irreducible multiplets as
\begin{equation}
8 \times 8 = 27 + 8s + 1 + 10^* + 10 + 8a
\end{equation}
where the first three are symmetric and the last three are anti-symmetric representations,
and the S-wave baryon-baryon interactions for example are reduced to only six independent interactions. 
Although the flavor {\it SU}(3) symmetry is approximate in the physical world,
the symmetry is useful in the physical world too and also in our lattice QCD simulation with the K-configuration set.
Therefore, let us see the baryon-baryon interactions expressed in the flavor irreducible representation basis.

\begin{figure}[t]
 \begin{tabular}{ccc}
\includegraphics[width=0.30\textwidth]{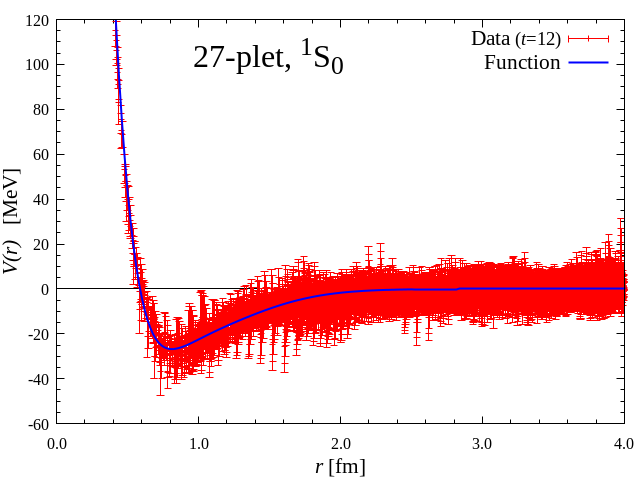}&
\includegraphics[width=0.30\textwidth]{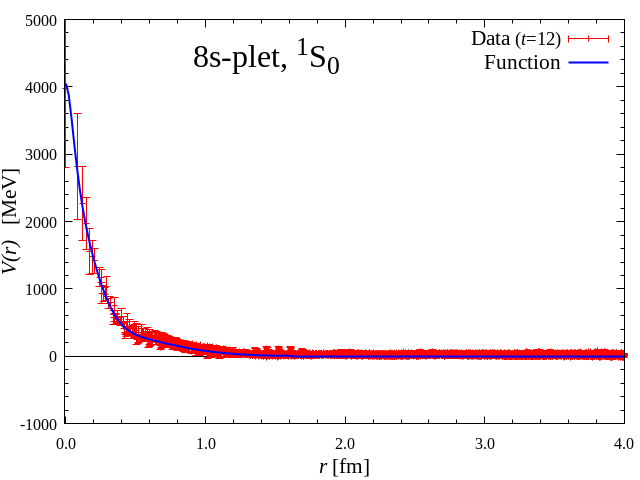}&
\includegraphics[width=0.30\textwidth]{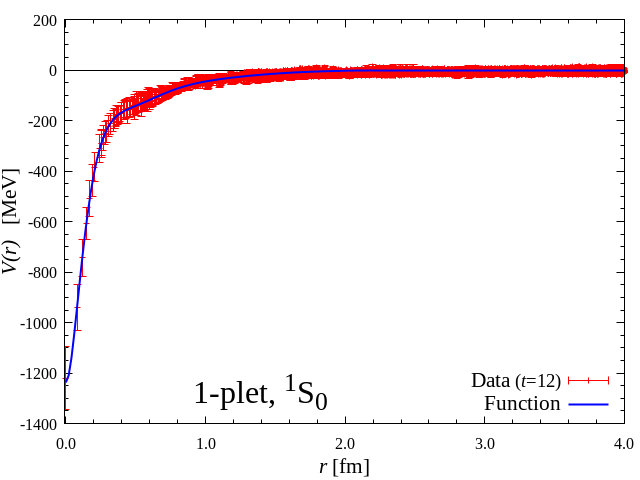}\\
\includegraphics[width=0.30\textwidth]{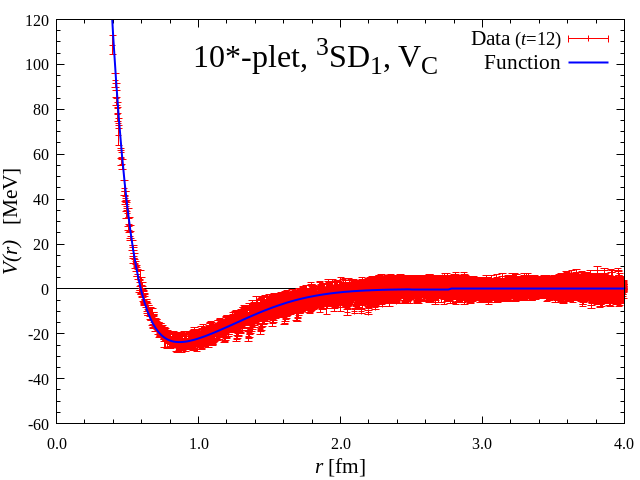}&
\includegraphics[width=0.30\textwidth]{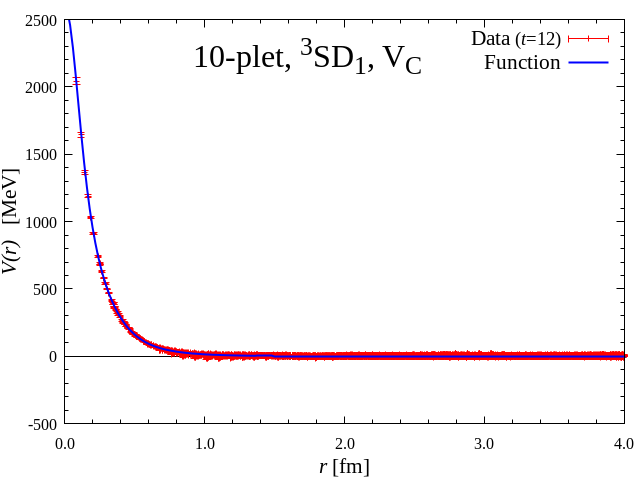}&
\includegraphics[width=0.30\textwidth]{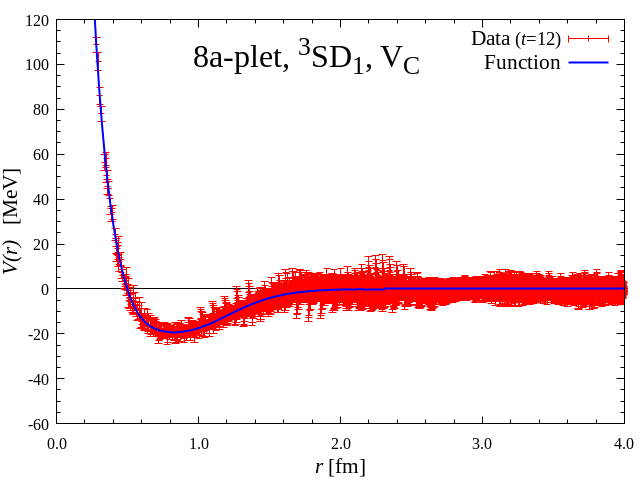}\\
\includegraphics[width=0.30\textwidth]{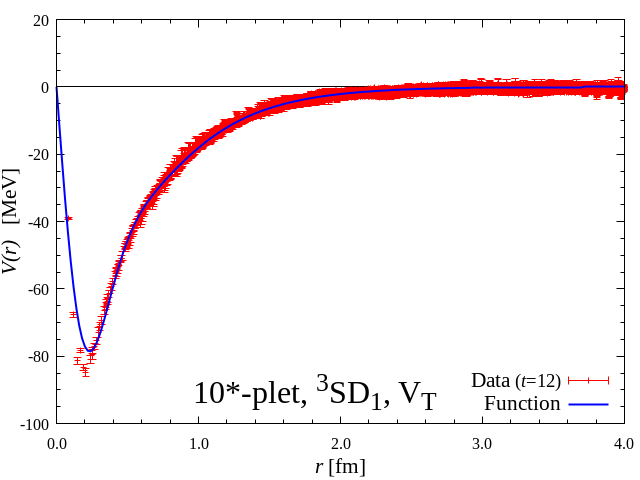}&
\includegraphics[width=0.30\textwidth]{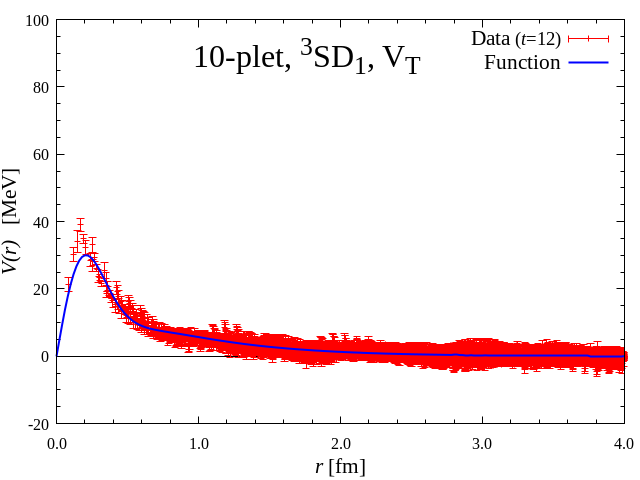}&
\includegraphics[width=0.30\textwidth]{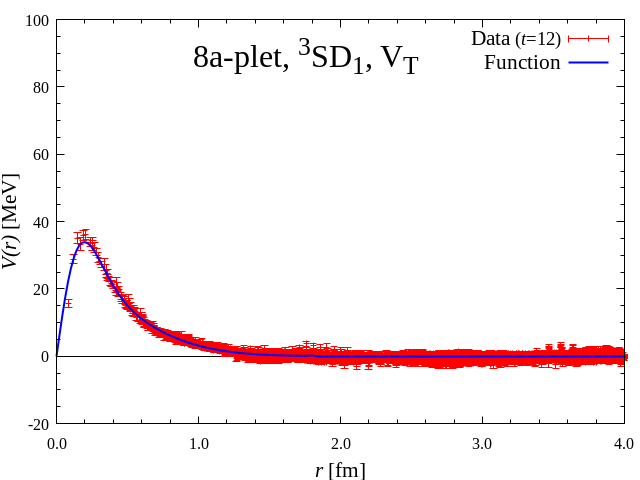}
 \end{tabular}
\caption{Potential of baryon-baryon S-wave interactions diagonal in the flavor irreducible representation basis.
These are obtained by rotating hyperon interaction potentials in the baryon-basis in strangeness $S$=$-2$ sector.}
\label{fig:potentials}
\end{figure} 

In Fig.~\ref{fig:isospin0}, one $3\times 3$ potential matrix in the baryon-basis is given
for the baryon-baryon interaction in $S$=$-2$, $I$=$0$ sector with $^1S_0$ partial wave.
We can rotate it into a matrix in the flavor-basis \{27,8s,1\}, 
by multiplying a $3\times 3$ matrix of the {\it SU}(3) Clebsch-Gordan coefficients as 
\begin{equation}
\left(
 \begin{array}{ccc}
  V_{ 27,27} &  V_{ 27,8s} & V_{ 27,\,1} \\
  V_{ 8s,27} &  V_{ 8s,8s} & V_{ 8s,\,1} \\
  V_{\,1,27} &  V_{\,1,8s} & V_{\,1,\,1}
 \end{array}   
\right)
=
A^t \left(
 \begin{array}{ccc}
  V_{\Lambda\Lambda,\Lambda\Lambda} &  V_{\Lambda\Lambda, N \Xi} & V_{\Lambda\Lambda, \Sigma\Sigma} \\
  V_{N \Xi,         \Lambda\Lambda} &  V_{N \Xi,          N \Xi} & V_{N \Xi,          \Sigma\Sigma} \\
  V_{\Sigma\Sigma,  \Lambda\Lambda} &  V_{\Sigma\Sigma,   N \Xi} & V_{\Sigma\Sigma,   \Sigma\Sigma}
 \end{array}   
\right) A
\quad \mbox{with} \quad
A =
\frac{1}{\sqrt{40}}
\left(
 \begin{array}{ccc}
    \sqrt{27} & -\sqrt{ 8} & -\sqrt{ 5} \\ 
    \sqrt{12} &  \sqrt{ 8} &  \sqrt{20} \\ 
           -1 & -\sqrt{24} &  \sqrt{15}
 \end{array}   
\right)
\end{equation}
where off-diagonal elements in the flavor-basis are allowed to exist because the flavor {\it SU}(3) symmetry is not exact in our QCD simulation.
Similarly, we can rotate the potential matrix in $S$=$-2$, $I$=$1$ sector with $^3S_1$-$^3D_1$ partial waves given in Fig.~\ref{fig:isospin1}, 
into a potential matrix in the flavor-basis \{10*,10,8a\}.

Figure~\ref{fig:potentials} shows diagonal parts of flavor-basis potential obtained,
where the upper three figures correspond to flavor-symmetric two-baryon in $^1S_0$ partial wave,
and the lower six figures correspond to flavor-anti-symmetric two-baryon in $^3S_1$-$^3D_1$ partial waves.
Feature of these diagonal potentials are identical with those observed at flavor {\it SU}(3) symmetric points~\cite{Inoue:2010hs}.
For example, entire attraction in the flavor singlet two-baryon and extreme repulsion in the flavor symmetric octet two-baryon are characteristic.
The potentials of 27-plet and 10*-plet are nothing but two-nucleon potential in the flavor {\it SU}(3) symmetric limit.
We see that a phenomenologically reasonable nuclear force is derived from QCD on lattice.
Moreover, we can see that a strong tensor component of nuclear force, which is also known phenomenologically, is reproduced from QCD.
These are important success of the HAL QCD method which had already been reported~\cite{Ishii:2006ec}.
Off diagonal components which we omit to show, are much weaker than the diagonal ones, 
since approximate flavor {\it SU}(3) symmetry holds well in our lattice QCD simulation as in the physical world.

As you can see in Fig.~\ref{fig:isospin0}, Fig.~\ref{fig:isospin1} and Fig.~\ref{fig:potentials}, 
data of lattice QCD induced potentials exist only at discrete distances.
Therefore, when we want to apply these potentials to investigate some physics, we need to parameterize them at first.
Blue curves in Fig.~\ref{fig:potentials} are plot of the parameterized potentials.
We have fitted following simple analytic functions to data in the least squares method,
\begin{eqnarray}
 V_{C}(r) &=& a_1 e^{-a_2\,r^2} + a_3 e^{-a_4\,r^2} + a_5 \left( (1 - e^{-a_6\,r^2}) \frac{e^{-a_7\,r}}{r} \right)^2
 \\
 V_{T}(r) &=& b_1 \left( 1 - e^{-b_2\,r^2} \right) \left(1 + \frac{3}{b_3\,r} + \frac{3}{(b_3\,r)^2} \right)\frac{e^{-b_3\,r}}{r} ~+~
              b_4 \left( 1 - e^{-b_5\,r^2} \right) \left(1 + \frac{3}{b_6\,r} + \frac{3}{(b_6\,r)^2} \right)\frac{e^{-b_6\,r}}{r}
\end{eqnarray}
for the central and tensor component, respectively.

\section{Single-particle potential of hyperons in nuclear matter}

\begin{table}[t]
\caption{Hyperon-nucleon coupled channels with a given total charge $Q$ and flavor symmetric $S$ or anti-symmetric $A$. 
In this study, $S$ ($A$) is combined with ${}^1S_0$ (${}^3S_1$-${}^3D_1$) partial wave.}
\label{tbl:channel}
 \begin{tabular}{l c c c c}
 \toprule
                 &  $Q=0$    &   $Q=+1$  &   $Q=-1$   & $Q=+2$ \\
  \midrule
  $(YN)_{S,A}$   & $\Lambda  n$, $\Sigma^0 n$, $\Sigma^- p$ & $\Lambda  p$, $\Sigma^0 p$, $\Sigma^+ n$ & $\Sigma^- n$ & $\Sigma^+ p$ \\
  \midrule
  $(\Xi N)_{S}$  & $\Xi^0 n$, $\Xi^- p$,  $\Sigma^+ \Sigma^-$, $\Sigma^0 \Sigma^0$, $\Sigma^0 \Lambda$, $\Lambda \Lambda$ &
                   $\Xi^0 p$, $\Sigma^+ \Lambda$  & $\Xi^- n$, $\Sigma^- \Lambda$ & \\
  \midrule
  $(\Xi N)_{A}$  & $\Xi^0 n$, $\Xi^- p$, $\Sigma^+ \Sigma^-$, $\Sigma^0 \Lambda$ &  $\Xi^0 p$, $\Sigma^+ \Sigma^0$, $\Sigma^+ \Lambda$ &
                   $\Xi^- n$, $\Sigma^- \Sigma^0$, $\Sigma^- \Lambda$  & \\
 \bottomrule
 \end{tabular}
\normalsize
\end{table}

In this section, let us study hyperons in nuclear matter by basing the hyperon forces induced from QCD.
Nuclear matter is a hypothetical uniform matter consists of infinite number of nucleon interacting each other via nuclear force.
Several theories have been developed to deal with the nucleonic matter.
Among them, we adopt the lowest order of the Brueckner theory in this paper, namely the Brueckner-Hartree-Fock (BHF) approximation.
In this frame work, single-particle potential of hyperon $Y$, 
in the nuclear matter with a total density $\rho=\rho_n + \rho_p$ and a proton fraction $x=\rho_p/\rho$,
is obtained by summing elements of the so called G-matrix.
\begin{equation}
 \label{eqn:Uy}
 U_Y(k,\rho,x) = \sum_{N=n,p} \, \sum_{^S\!L_J} \, \sum_{k'\leq k_F^{(N)}}
                  \mbox{Re}~\mate<k,k'| G^{^S\!L_J}_{YN,YN}(E_{YN}(k,k')) |k,k'>
\end{equation}
The G-matrix $G^{^S\!L_J}_{YN,YN}(\omega)$ describes $YN$ to $YN$ scattering in the nucleonic matter in the partial wave ${}^{S}\!L_{J}$,
and can be obtained as a solution of the Bethe-Goldstone equation with a interaction potential $V_{a,b}$
\begin{equation}
 \label{eqn:BGeq}
     G_{a,b}(\omega) = V_{a,b} + \sum_c \sum_{k,k'}
                       V_{a,c} \, \ketv k,k'> \frac{Q_c(k,k')}{\omega-E_c(k,k')+i\epsilon} \brav k,k'| \, G_{c,b}(\omega)
\end{equation}
where $Q(k,k')$ is the angle averaged Pauli operator, and $E(k,k')$ is the energy of intermediate baryon pair given by
\begin{equation}
 \label{eqn:Ene}
  E_{BB'}(k,k') = e_B(k,\rho,x) + e_{B'}(k',\rho,x) \quad \mbox{and} \quad  e_B(k,\rho,x) = M_B + \frac{k^2}{2 M_B} + U_B(k,\rho,x)
\end{equation}
in the so called continuous choice. 
The last equation is the definition of single-particle potential of baryon in nuclear matter $U_B(k,\rho,x)$,
which is a part of energy spectrum of baryon in nuclear matter $e_B(k,\rho,x)$.

These highly coupled equations are solved in the iteration procedure,
and self-consistent G-matrix and potential $U_B(k,\rho,x)$ are obtained.
In this paper, we consider two representative nuclear matter, namely, the pure neutron matter (PNM) with the proton fraction $x=0$,
and the symmetric nuclear matter (SNM) with $x=1/2$.
In addition, we fix density of matter at the so called normal nuclear matter density $\rho_0=0.17$~fm$^{-3}$.
Therefore, we suppress arguments $\rho$ and $x$ of $U_B$ from here.

We need to proceed step by step, because calculation of $U_{\Xi}(k)$ require $U_{\Lambda}(k)$ and $U_{\Sigma}(k)$ determined,
and calculation of $U_{\Lambda}(k)$ and $U_{\Sigma}(k)$ require $U_N(k)$ the single-particle potential of nucleon determined.
We use $U_p(k)$ and $U_n(k)$ obtained in our BHF calculation by using the AV18 phenomenological two-nucleon force~\cite{Wiringa:1994wb}
supplemented by the Urbana type three-nucleon force~\cite{Carlson:1983kq}.
It turns out that resulting $U_Y(k)$ are not sensitive to $U_p(k)$ and $U_n(k)$ used.
For baryon mass $M_B$, we use realistic values $M_N$=$939$ MeV, $M_{\Lambda}$=$1116$ MeV, $M_{\Sigma}$=$1193$ MeV and $M_{\Xi}$=$1318$ MeV,
because baryon masses obtained in the present lattice QCD simulation are almost realistic,
and we want to focus on the effect of hyperon interactions theoretically obtained.

Note that hyperon-nucleon interactions connect many two-baryon channels.
Table~\ref{tbl:channel} list the coupled channels which enter in this calculation.
We do not ignore any interaction in channels but include all, for example, $\Lambda\Lambda$ and $\Lambda\Sigma$ interactions are included in our calculation.
While, the sum about partial waves in eq.(\ref{eqn:Uy}) is restricted to ${}^1S_0$ and ${}^3S_1$-${}^3D_1$ in this paper, 
since our lattice QCD induced hyperon interaction is available only in these partial waves at present.
This truncation will be reasonable for low density nucleonic matter. We discuss this point later.

Since this study is our first attempt to apply hyperon interaction derived from QCD to strange nuclear physics,
let us begin with hyperon interaction potentials reconstructed from
the flavor-basis diagonal components and the {\it SU}(3) Clebsch-Gordan coefficients, for both $S$=$-1$ and $S$=$-2$ sectors.
In this case, we need only the nine parameterized potentials shown in Fig.~\ref{fig:potentials}.

\begin{figure}[t]
\centering
\includegraphics[width=0.4\textwidth]{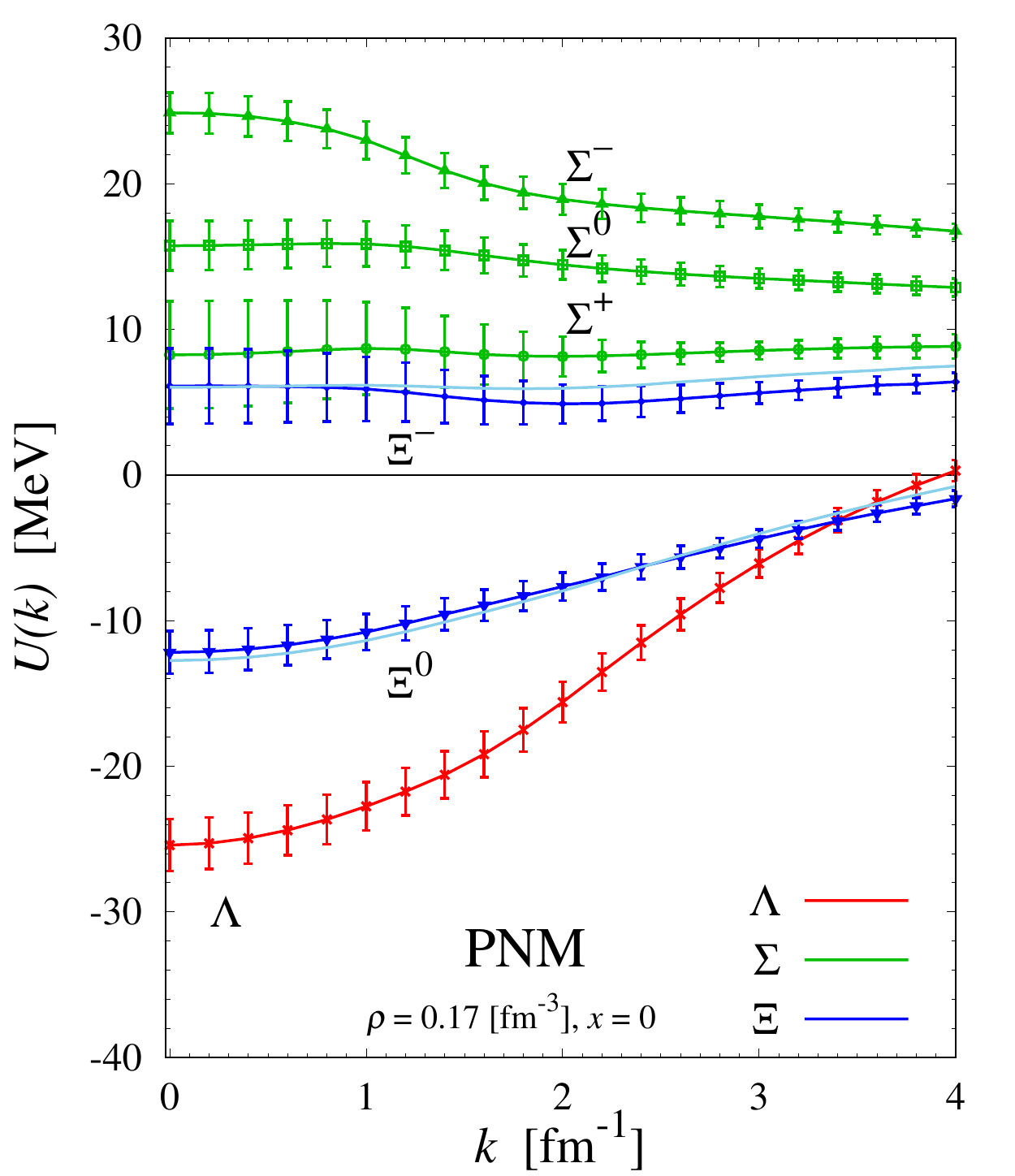} \qquad
\includegraphics[width=0.4\textwidth]{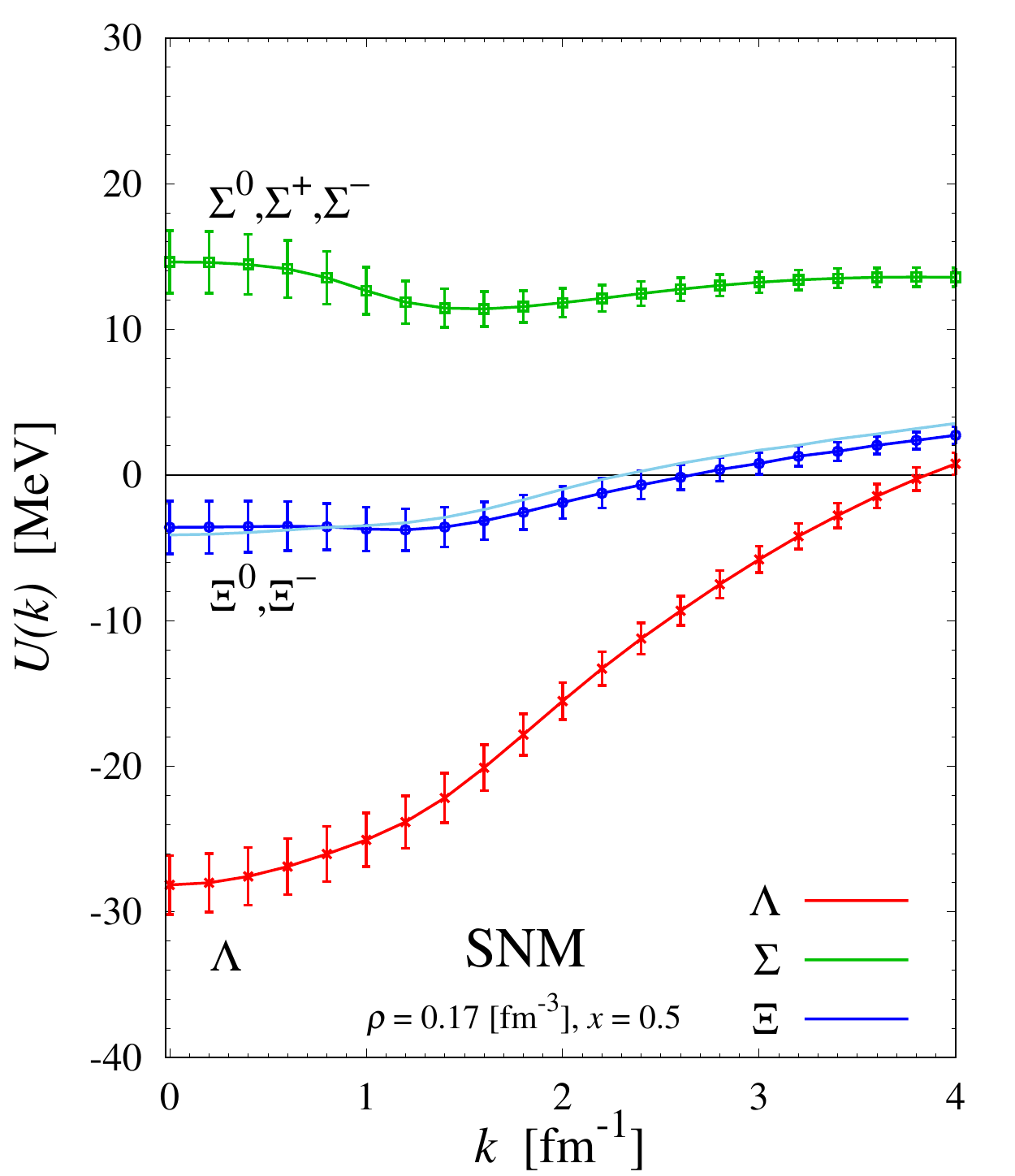}
\caption{Hyperon single-particle potentials $U_Y(k)$ in nucleonic matter with the normal nuclear density, 
based on the hyperon interaction potentials from QCD on lattice.}
\label{fig:uys}
\end{figure} 

Figure~\ref{fig:uys} shows obtained single-particle potential $U_Y(k)$ in PNM at the left panel and SNM at the right panel, 
as a function of hyperon momentum $k$ against the nuclear matter.
The vertical bars stand for error associated with the statistical error in our QCD induced potentials, 
which are also estimated with the Jackknife method.
We see that combination of QCD and the traditional many-body theory predict that, 
for hyperons stopping in SNM at the normal nuclear matter density,
$U_{\Lambda}(0)=-28$~MeV, $U_{\Sigma}(0)=+15$~MeV, and $U_{\Xi}(0)=-4$~MeV,
as the central value with the statistical error about $\pm 2$ MeV.

Empirical value of $U_Y(0)$ in SNM have been estimated through hypernuclear experiment,
since matter at center of heavy nuclei is analogous to SNM.
At present, experimental data indicate that 
$U_{\Lambda}(0) \simeq -30$~MeV, $U_{\Sigma}(0) \geq 20$~MeV and  $U_{\Xi}(0) \simeq -10$~MeV.
Because there is rich data of $\Lambda$-hypernuclei, the empirical value of $U_{\Lambda}(0)$ is rather robust.
While, experimental data on $\Sigma$ and $\Xi$ in nuclei are very limited,
hence the above $U_{\Sigma}(0)$ and $U_{\Xi}(0)$ should be confirmed and refined in further experiment.

We see that our theoretical result of $U_Y(0)$ in SNM agree with the experimental ones.
The overall agreement is excellent and remarkable.
Recall that we have not used any phenomenological input for hyperon interactions but used only QCD.
Therefore, we claim that this success is the first-ever explanation of strange nuclear physics starting from QCD.
Moreover, we consider that the above less certain experimental indication of $U_{\Sigma}(0)$ and $U_{\Xi}(0)$ 
are qualitatively supported by QCD. 
Especially, attraction of $\Xi$ in SNM is important since it support existence of $\Xi$ hypernucleus.

\begin{table}[t]
\caption{Breakdown of the potential $U_{Y}(0)$ in SNM at $\rho_0$ into the G-matrix contribution,
where both the spin and isospin multiplicities are included.}
\label{tbl:breakdown1}
\begin{tabular}{cccccccc}
\toprule
\multirow{3}{*}{$\Lambda$}   &           &   $I=1/2$   &           &            &             &           & \multirow{2}{*}{Total [MeV]} \\
                             &  $^1S_0$  &  $^3S_1$    &  $^3D_1$  &            &             &           &                        \\
\cmidrule(rl){2-4}
\cmidrule(l){8-8}
                             &  $-3.49$  &  $-24.84$   &  $0.18$   &            &             &           & $-28.16$               \\
\bottomrule
\toprule
\multirow{3}{*}{$\Sigma$}    &           &   $I=1/2$   &           &            &   $I=3/2$   &           & \multirow{2}{*}{Total [MeV]} \\
                             &  $^1S_0$  &  $^3S_1$    &  $^3D_1$  &  $^1S_0$   &  $^3S_1$    &  $^3D_1$  &                        \\
\cmidrule(rl){2-4}
\cmidrule(rl){5-7}
\cmidrule(l){8-8}
                             &   $7.43$  &   $-9.28$   &  $0.07$   &  $-4.97$   &   $21.80$   &  $-0.43$  &  $14.62$               \\
\bottomrule
\toprule
\multirow{3}{*}{$\Xi$}       &           &   $I=0$     &           &            &   $I=1$     &           & \multirow{2}{*}{Total [MeV]} \\
                             &  $^1S_0$  &  $^3S_1$    &  $^3D_1$  &  $^1S_0$   &  $^3S_1$    &  $^3D_1$  &                        \\
\cmidrule(rl){2-4}
\cmidrule(rl){5-7}
\cmidrule(l){8-8}
                             &  $-4.48$  &  $-4.37$    &  $-0.01$  &  $9.08$    &   $-3.74$   &  $-0.08$  &  $-3.60$               \\
\bottomrule
\end{tabular}
\end{table}

Table~\ref{tbl:breakdown1} lists breakdown of the potential $U_{Y}(0)$ in SNM into the G-matrix contributions.
Note that both the spin and isospin multiplicities are included.
It is interesting to compare these values to those based on various models of hyperon interaction.
For example, the dominant attractive contribution to $U_{\Lambda}(0)$ from interaction in $^3S_1$ partial wave,
is characteristic compared to other theoretical studies~\cite{Baldo:1999rq,Kohno:2009vk,Yamamoto:2014jga}.
We see that $\Sigma$ receive a repulsion in SNM because the repulsion from \{$I$=$1/2$, $^1S_0$\} and \{$I$=$3/2$, $^3S_1$\}
is stronger than the attraction from \{$I$=$1/2$, $^3S_1$\} and \{$I$=$3/2$, $^1S_0$\}.
We see that the attraction of $\Xi$ in SNM is small because the repulsion in \{$I$=$1$,$^1S_0$\}
almost cancel out the attraction from other channels.

Large part of neutron star core basically consist of the pure neutron matter,
though proton, electron and muon mix in it a little depending on the density for the beta equilibrium and charge neutrality.
Hyperon will emerge in the matter depending on chemical potential of hyperon $\mu_Y$ compared to that of neutron.
Since $\mu_Y$ is given by $\mu_{Y}=M_{Y}+U_{Y}(0)$ in good approximation,
theoretical prediction of $U_Y(k)$ in PNM is important to study neutron stars and very interesting.
In Fig.~\ref{fig:uys}, we see that overall feature of resulting $U_Y(k)$ in PNM is similar to that of SNM,
namely, $\Lambda$ receive an attraction about 30~MeV and $\Sigma$ receive a repulsion about 20~MeV at the density $\rho_0$.
One remarkable point is that the potential of $\Xi$ in PNM strongly depends on change of $\Xi$,
so that $\Xi^0$ receive an attraction but $\Xi^-$ receive a repulsion in PNM.
This suggest that more isospin symmetric nuclei are more advantageous to bind $\Xi^-$.
It is interesting to confirm this in future experiment.

For hyperon interactions in $S$=$-2$ sector, the original lattice QCD induced potentials
are given in Fig.~\ref{fig:isospin0} and Fig.~\ref{fig:isospin1}.
We can use them in calculation of $U_{\Xi}(k)$, though we need to parameterize all these potentials.
While, our lattice QCD potentials in $S$=$-1$ sector are less clear than $S$=$-2$ sector~\cite{Nemura:2018},
and not easy to use in this application at present, unfortunately.
Therefore, let us calculate only $U_{\Xi}(k)$ by using the original $S$=$-2$ potentials including the explicit flavor {\it SU}(3) breaking, 
and $U_{\Lambda}(k)$ and $U_{\Sigma}(k)$ obtained with the reconstructed $S$=$-1$ potentials.

Sky-blue curves without bars in Fig.~\ref{fig:uys} show the resulting $U_{\Xi}(k)$,
where only central value are shown without statistical error associated with our Monte Carlo simulation.
We see that sky-blue curves almost agree with blue ones which are $U_{\Xi}(k)$ obtained
with the approximately reconstructed $S$=$-2$ potentials.
This indicate that our approximation reconstructing hyperon potentials from the flavor-basis diagonal components is reasonable,
and we can regard resulting $U_{Y}(k)$ are qualitatively correct.
Small difference between blue and sky-blue curve shows effect of the explicit flavor {\it SU}(3) breaking.
We expect similar size of the effect for $U_{\Lambda}(k)$ and $U_{\Sigma}(k)$.
Table~\ref{tbl:breakdown2} lists breakdown of $U_{\Xi}(0)$ in SNM with the original potentials into the G-matrix contributions.
By comparing Table~\ref{tbl:breakdown1} and Table~\ref{tbl:breakdown2},
we see that some effect of the approximation exist in each contribution.
This means that we should apply the original potentials in $S$=$-1$ sector
when we study $U_{\Lambda}(k)$ and $U_{\Sigma}(k)$ quantitatively precisely.
This is one of our future plans.

\section{Summary and outlook}

In this paper, we've studied $\Lambda$, $\Sigma$, and $\Xi$ hyperons in nuclear matter starting from QCD on lattice.
First, we've derived potential of hyperon interactions from QCD by means of the HAL QCD method. 
Then, we've applied obtained potentials to the BHF many-nucleon theory,
and calculate single-particle potential of hyperon in nuclear matter $U_Y(k)$.
The obtained $U_Y(0)$ in SNM at $\rho_0$ remarkably agree with the current experimental information.
We believe that this achievement is a significant progress in strange nuclear physics. 

We've begun with the hyperon interaction potentials reconstructed from the flavor-basis diagonal potentials.
Then, we've replaced the potentials in $S$=$-2$ sector to the original ones.
We've found that the approximation is reasonable for qualitative study.
We plan to apply the original hyperon potentials in $S$=$-1$ sector, so that we can include effect
of the physical explicit flavor {\it SU}(3) breaking to $U_{\Lambda}(k)$ and $U_{\Sigma}(k)$ for quantitative study.

One may be interested in whether the HAL QCD $\Lambda N$-$\Sigma N$ potentials explain rich data of $\Lambda$-hypernuclei.
This is what we are trying to clarify.
We cannot expect perfect reproduction of whole data at this moment, since we have only S-wave potentials,
and quark masses in our QCD simulation are slightly different from that of the real world.
However, we expect that we can reproduce bulk of data at least qualitatively,
since the present result of $U_{\Lambda}(0)$ and $U_{\Sigma}(0)$ in SNM agree with experimental ones as we have seen.

In our calculation of $U_Y(k)$, we've truncated the partial wave expansion of G-matrix
at $^3S_1$-$^3D_1$ due to absence of QCD induced hyperon interactions at present.
This truncation must be reasonable at $\rho_0$ and below. 
While, at deep inside of neutron star core, density of matter becomes much higher than $\rho_0$. 
Hence, if we want to study hyperon emergence in neutron star core, we must calculate $U_Y(k)$ at high density.
Since there are high momentum nucleons in high density matter, 
hyperon-nucleon interaction in higher partial waves will contribute to $U_Y(K)$ sizably, and we need to include them.
Especially, we need to include P-wave hyperon-nucleon interactions first.
We've already developed a technique to extract two-baryon interactions in odd parity partial waves from QCD on lattice,
and applied it at large quark mass~\cite{Ishii:2014lra}.
We could reveal flavor-spin nature of P-wave two-baryon force including symmetric and anti-symmetric spin-orbit forces.
It also turned out that computational cost to extract P-wave two-baryon force is enormously large compared to S-wave force,
and the extraction at the physical point is impractical within the ability of today's large scale computer.
Fortunately, construction of a new large scale computer is ongoing in Japan.
We will be able to extract $YN$ and $YY$ P-wave force from QCD at the physical point, 
and predict $U_Y(k)$ at high density, in near future.
In addition, we hope we can derive also hyperon three-body force from QCD 
by using the new large scale computer, which may be important in high density matter.

\begin{table}[t]
\caption{Breakdown of the potential $U_{\Xi}(0)$ in SNM at $\rho_0$ calculated with
the original QCD $S$=$-2$ two-baryon potentials including the flavor {\it SU}(3) breaking.}
\label{tbl:breakdown2}
\centering
\begin{tabular}{cccccccc}
\toprule
\multirow{3}{*}{$\Xi$}       &           &  ~~$I=0$~~  &           &            &  ~~$I=1$~~  &           & \multirow{2}{*}{Total [MeV]} \\
                             &  $^1S_0$  &  $^3S_1$    &  $^3D_1$  &  $^1S_0$   &  $^3S_1$    &  $^3D_1$  &                        \\
\cmidrule(rl){2-4}
\cmidrule(rl){5-7}
\cmidrule(l){8-8}
                             &  $-3.15$  &  $-5.36$    &  $-0.30$  &  $7.12$    &   $-2.41$   &  $-0.08$  &  $-4.11$           \\
\bottomrule
\end{tabular}
\end{table}

\section{ACKNOWLEDGMENTS}
We thank the PACS Collaboration for generating and providing gauge configurations,
and the JLDG team~\cite{JLDG,Amagasa:2015zwb} for providing storage to save our data. 
Numerical computation of this work was carried out on 
the K computer at RIKEN R-CCS (hp120281, hp130023, hp140209, hp150223, hp150262, hp160211, hp170230), 
the HOKUSAI FX100 computer at RIKEN Wako (G15023, G16030, G17002),
and the HA-PACS at University of Tsukuba (14a-20, 15a-30).
This research is supported in part by the JSPS Grant-in-Aid for Scientific Research (C)18K03628,
Strategic Program for Innovative Research (SPIRE) Field 5 project, 
"Priority Issue on Post-K computer" (Elucidation of the Fundamental Laws and Evolution of the Universe),
and Joint Institute for Computational Fundamental Science (JICFuS).



\begin{thebibliography}{99}
\bibitem{Nemura:2018} 
  H.~Nemura for HAL QCD Collaboration, will be published in this volume
\end{thebibliography}
\end{document}